\documentclass[aps,prb,twocolumn,superscriptaddress,notitlepage]{revtex4-1}
\usepackage{graphicx}
\usepackage[caption=false]{subfig}
\usepackage{float}
\usepackage{natbib}
\usepackage{xcolor}
\usepackage{xr}
\usepackage{amsmath}
\usepackage{amssymb}
\usepackage{array}
\usepackage{wasysym}
\usepackage{color,soul}
\usepackage{braket}
\usepackage{verbatim}
\usepackage{hyperref}
\usepackage{gensymb}
\usepackage{url}
\usepackage{dirtytalk}

\usepackage{filecontents}
\begin{document}

\title{Competing Quantum Orders in 6R-TaS$_2$: Unconventional Superconductivity, Charge Order, and an Anomalous Hall Effect phase}

\author{V. Sazgari}
\affiliation{PSI Center for Neutron and Muon Sciences CNM, 5232 Villigen PSI, Switzerland}

\author{J.N. Graham}
\affiliation{PSI Center for Neutron and Muon Sciences CNM, 5232 Villigen PSI, Switzerland}

\author{S.S. Islam}
\affiliation{PSI Center for Neutron and Muon Sciences CNM, 5232 Villigen PSI, Switzerland}

\author{P. Král}
\affiliation{PSI Center for Neutron and Muon Sciences CNM, 5232 Villigen PSI, Switzerland}

\author{O. Gerguri}
\affiliation{PSI Center for Neutron and Muon Sciences CNM, 5232 Villigen PSI, Switzerland}

\author{A. Achari}
\affiliation{National Graphene Institute, University of Manchester, Manchester M13 9PL, United Kingdom}
\affiliation{Department of Chemical Engineering, University of Manchester, Manchester M13 9PL, United Kingdom}

\author{J.N. Tangermann}
\affiliation{Department of Quantum Matter Physics, University of Geneva, CH-1211 Geneva, Switzerland}

\author{H. Gopakumar}
\affiliation{PSI Center for Neutron and Muon Sciences CNM, 5232 Villigen PSI, Switzerland}

\author{G. Simutis}
\affiliation{PSI Center for Neutron and Muon Sciences CNM, 5232 Villigen PSI, Switzerland}

\author{M.~Janoschek}
\affiliation{PSI Center for Neutron and Muon Sciences CNM, 5232 Villigen PSI, Switzerland}
\affiliation{Physik-Institut, Universität Zurich, Winterthurerstrasse 190, CH-8057 Zurich, Switzerland.}

\author{M. Bartkowiak}
\affiliation{PSI Center for Neutron and Muon Sciences CNM, 5232 Villigen PSI, Switzerland}

\author{R. Khasanov}
\affiliation{PSI Center for Neutron and Muon Sciences CNM, 5232 Villigen PSI, Switzerland}

\author{H. Luetkens}
\affiliation{PSI Center for Neutron and Muon Sciences CNM, 5232 Villigen PSI, Switzerland}

\author{F.O.~von~Rohr}
\affiliation{Department of Quantum Matter Physics, University of Geneva, CH-1211 Geneva, Switzerland}

\author{R.R. Nair}
\affiliation{National Graphene Institute, University of Manchester, Manchester M13 9PL, United Kingdom}
\affiliation{Department of Chemical Engineering, University of Manchester, Manchester M13 9PL, United Kingdom}

\author{Z. Guguchia}
\email{zurab.guguchia@psi.ch}
\affiliation{PSI Center for Neutron and Muon Sciences CNM, 5232 Villigen PSI, Switzerland}

\date{\today}


\maketitle

\textbf{The transition metal dichalcogenide 6R-TaS$_{2}$ offers a natural platform for studying the interplay among charge density wave (CDW) order, superconductivity, and transport anomalies. Recent findings reveal that, in the intermediate temperature range between charge order and superconductivity, a hidden order emerges around $T^{*}$ ${\simeq}$ 35 K-as evidenced by strong magnetoresistance and an anomalous Hall effect (AHE). However, the nature of the superconducting pairing, the hidden order, and their relationship with the CDW remain unclear. Using ${\mu}$SR, magnetotransport, and hydrostatic pressure techniques, we identify a nodal superconducting state with low superfluid density at ambient pressure, with no spontaneous magnetic order detected below $T^{*}$. This indicates that the AHE originates from the band structure rather than magnetism. Under pressures up to 2 GPa, the superfluid density rises markedly in correlation with the superconducting transition temperature, the nodal pairing shifts to a nodeless state, and the CDW onset is reduced by half. Notably, AHE is fully suppressed and magnetoresistance drops by 50${\%}$ within just 0.2 GPa, highlighting the fragility of the hidden order. These results reveal an unconventional superconducting pairing in 6R-TaS$_{2}$, competing with both CDW and hidden orders through weakened interlayer coupling and competition for the same electronic states. With a multifaceted approach, we establish a comprehensive phase diagram that reveals the intricate interplay and competition between the intertwined quantum orders in 6R-TaS$_{2}$.}

Understanding the mechanism of superconductivity and its interplay with various quantum states—including charge density wave (CDW) order, magnetism, and electronic responses such as the anomalous Hall effect (AHE), magnetoresistance (MR)-is a central topic in condensed matter physics \cite{keimer2015quantum,doi:10.1126/science.237.4819.1133,nagaosa2010anomalous,RevModPhys.56.755,Fernandes_2022}. The search for suitable material platforms where these intertwined quantum phases can be systematically explored remains an ongoing challenge. Among the most promising candidates are transition metal dichalcogenides (TMDs) \cite{novoselov20162d,manzeli20172d}, a diverse family of two-dimensional materials that exhibit a wide range of electronic properties, including semiconducting behavior, unconventional superconductivity \cite{guguchia2017signatures,doi:10.1126/sciadv.aav8465,islam2024pressure,Navarro_Moratalla_2016}, CDW states \cite{doi:10.1073/pnas.1512092112}, magnetism \cite{doi:10.1126/sciadv.aat3672,regmi2024altermagnetismlayeredintercalatedtransition,wang2022magnetic}, and topological states \cite{C6NR07851F,PhysRevB.108.245131,hsu2017topological,bae2025designingflatbandslocalized,PhysRevX.15.011049}.\\

	\begin{figure*}[t!]
		\centering
		\includegraphics[width=1.3\linewidth]{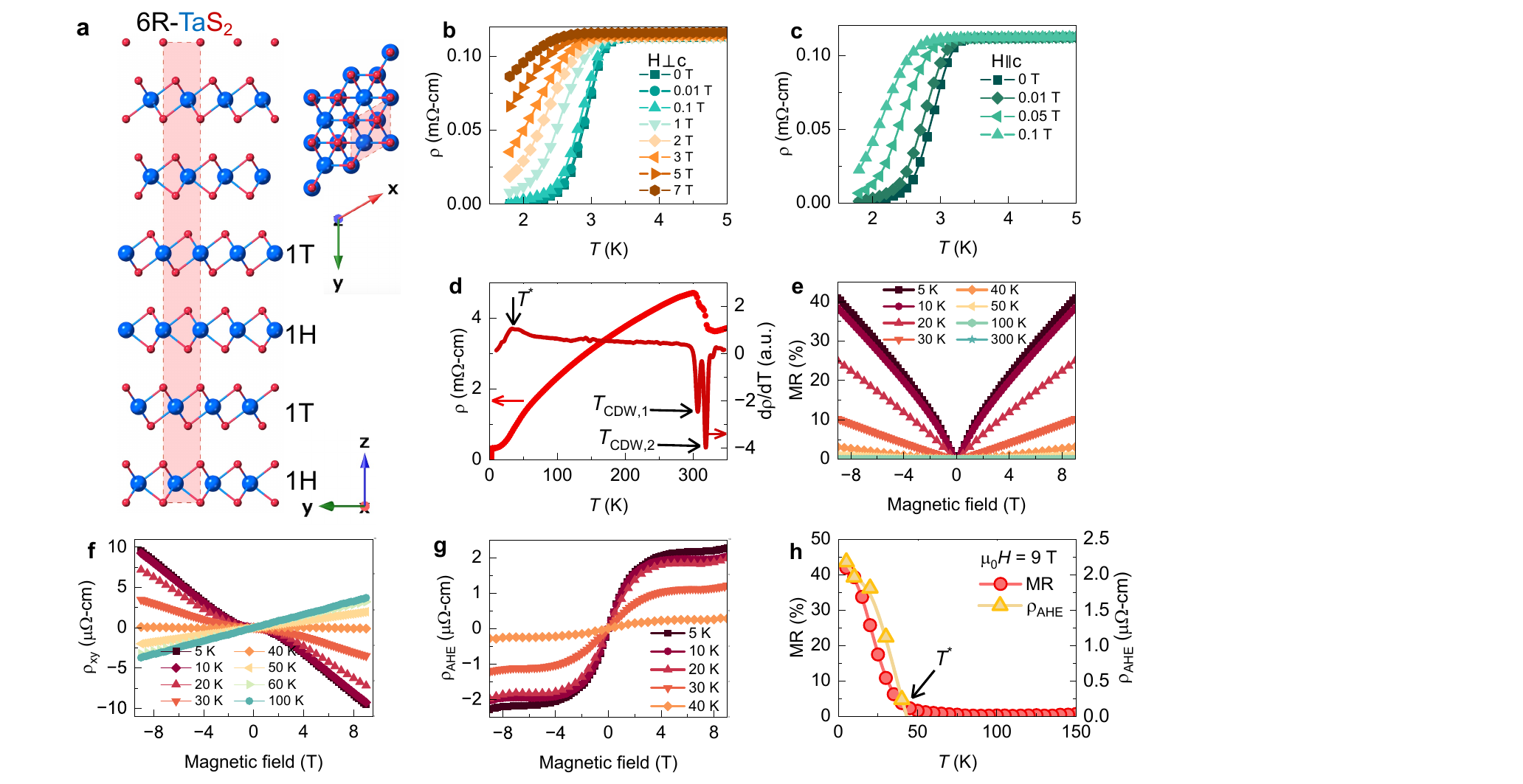}
		\vspace{-0.7cm}
		\caption{\textbf{Summary of transport results under ambient pressure.}
		$\bf{a,}$ Crystal structure of 6R-TaS$_{2}$ in two different orientations. The shaded region represents a unit cell. $\bf{b-c,}$ The temperature dependence of the electrical resistivity of a single crystal of 6R-TaS$_{2}$, focusing on the low-temperature region, measured under various magnetic fields applied perpendicular to the $c$-axis ($\bf{b}$) and parallel to the $c$-axis ($\bf{c}$). $\bf{d,}$ The temperature dependence of resistivity and its first derivative, measured in a wide temperature range. Arrows mark three characteristic temperatures $T_{\rm CDW,1}$, $T_{\rm CDW,2}$ and $T^{*}$. $\bf{e,}$ The magnetoresistance measured at various temperatures. $\bf{f,}$ The Hall resistance, measured at various temperatures between 5 K and 100 K. $\bf{g,}$ Field dependent anomalous Hall resistance $\rho_{\rm AHE}$, measured at various temperatures. $\bf{h,}$ The temperature dependence of the value of magnetoresistance (left axis) and the anomalous Hall resistance at 9 T (right axis).}
		\label{Fig1}
	\end{figure*}

Tantalum disulfide (TaS$_{2}$) is one of the most intriguing members of the TMD family \cite{achari2022alternating,pal2023charge,kvashnin2020coexistence,liu2024nematic,law20171t,ribak2020chiral}. In particular, the 6R-TaS$_{2}$ polymorph features a unique bulk layered heterostructure, consisting of alternating trigonal prismatically coordinated (1H) and octahedrally coordinated (1T) layers in the 6R phase (Fig. 1a). Recent studies have revealed that 6R-TaS$_{2}$ undergoes a CDW transition with $T_{\rm CDW}$ ${\simeq}$ 305 K, followed by a bulk superconducting transition at approximately $T_{\rm c}$ ${\simeq}$ 2.6 K \cite{achari2022alternating}, determined from magnetization measurements. It has been proposed that superconductivity primarily resides in the 1H layers, which are separated by insulating 1T layers in the CDW state, with minimal interference between the two electronic states. 
This unique alternating layered structure therefore positions 6R-TaS$_{2}$ as a potential host for 2D superconductivity in bulk form. The fact that $T_{\rm c}$ in 6R-TaS$_{2}$ exceeds that of both the 2H (0.8 K) and 1T phases (1.5 K at 2.5 GPa), further emphasizing the role of its distinct heterostructural nature. Additionally, 6R-TaS$_{2}$ exhibits an AHE and large MR below $T^{*}$ ${\simeq}$ 30 K \cite{liu2024nematic}, accompanied by features such as Kondo screening and nematic magnetoresistance. This was interpreted as an indication of the presence of hidden magnetism. High-pressure experiments reveal that superconductivity and charge order exhibit antagonistic pressure dependencies \cite{PhysRevLett.133.056001,Lv2024ReemergenceOS,doi:10.1021/acs.nanolett.4c00579}.

These findings thus reveal a complex interplay between CDW order, superconductivity, and magnetic phenomena, establishing 6R-TaS$_{2}$ as an ideal platform for exploring the coupling between these collective quantum phases and their influence on transport properties. Furthermore, this intrinsic stacking configuration offers a unique opportunity to investigate these interactions in a naturally occurring heterostructure. Despite the above findings, several fundamental questions about 6R-TaS$_{2}$ remain unanswered. For example, the microscopic nature of superconductivity, including the gap symmetry and superfluid density, has yet to be determined. Moreover, the precise origin of the transition at $T^{*}$, whether magnetic or not, remains unclear. Additionally, the relationship between superconductivity and the hidden state below $T^{*}$ remains elusive. In particular, it is uncertain if any competition between superconductivity and charge order arises solely from interlayer frustration or from a contest for the same electronic states.\\

	\begin{figure*}[t!]
		\centering
		\includegraphics[width=\linewidth]{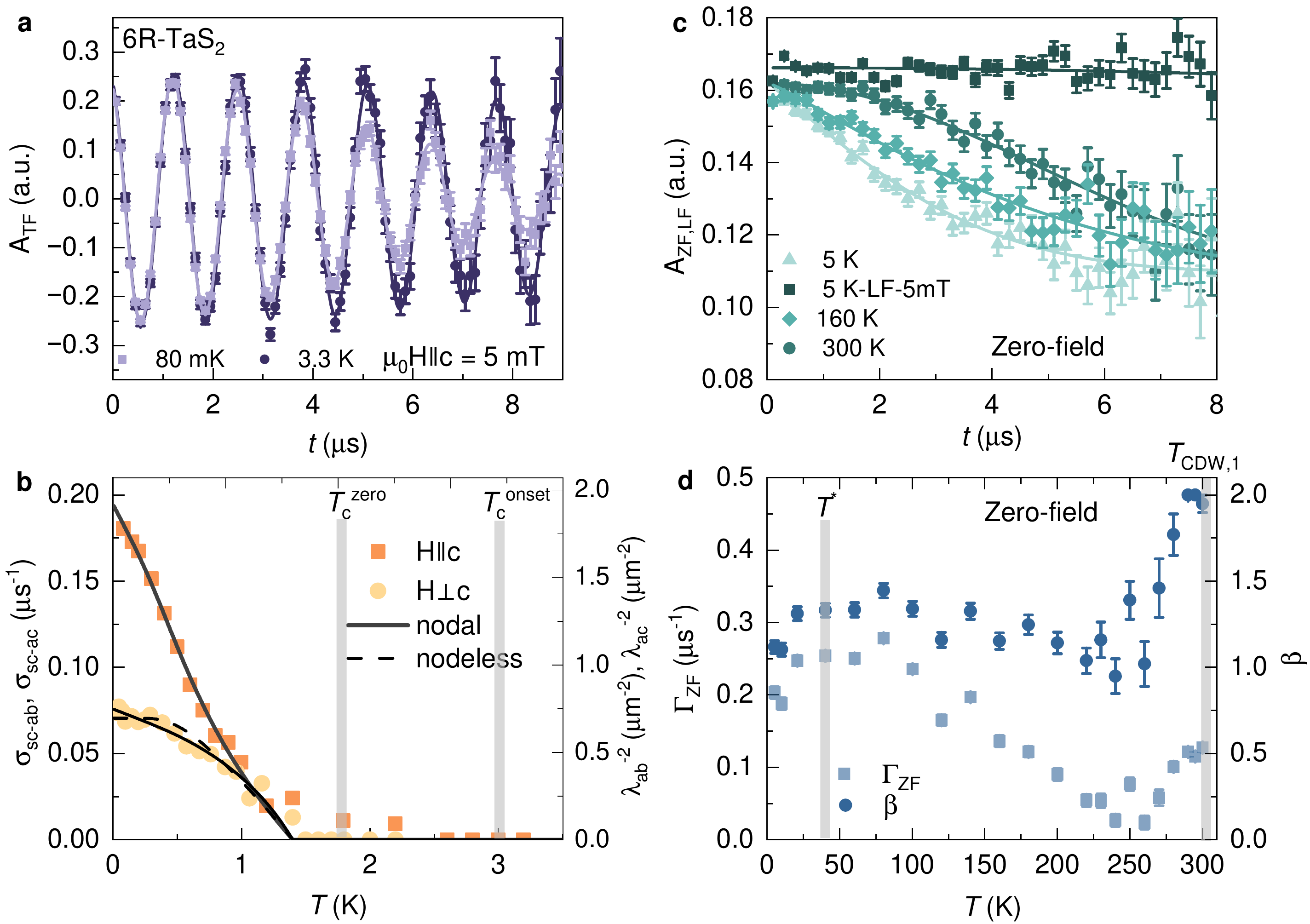}
		\vspace{-0.7cm}
		\caption{\textbf{Summary of ${\mu}$SR results at ambient pressure.}
		$\bf{a,}$ TF-$\mu$SR time spectra recorded above and below $T_{\rm c}$ after field cooling the sample from above $T_{\rm c}$ under an $c$-axis applied magnetic field of 5~mT for the single crystal of 6R-TaS$_{2}$. Solid lines correspond to the fits as described in the text. $\bf{b,}$ The ZF-$\mu$SR time
        spectra for 6R-TaS$_{2}$, recorded at various temperatures. The solid lines represent fits to the data by means of Eq. 3.  $\bf{c,}$ The temperature dependence of the superconducting muon spin depolarization rates ${\sigma}_{\rm SC-ab}$ and ${\sigma}_{\rm SC-ac}$ for 6R-TaS$_{2}$, measured in an applied magnetic field of 5 mT. The error bars represent the standard deviations of the fit parameters. The dashed (solid) lines correspond to a fit using a model with nodeless (nodal) gap superconductivity. Right axis shows the inverse squared effective penetration depths $\lambda_{ab}^{-2}$ and $\lambda_{ac}^{-2}$. Vertical grey lines mark the onset temperature of superconductivity and the temperature corresponding to the zero-resistance state, as determined from resistivity experiments. $\bf{d,}$ The temperature dependence of zero-field muon-spin relaxation rate ${\Gamma}_{\rm ZF}$ and the stretching exponent ${\beta}$.}
		\label{Fig1}
	\end{figure*}

To address these open questions, this work presents a comprehensive investigation of 6R-TaS$_{2}$, combining local probe techniques such as muon spin rotation (${\mu}$SR), magnetotransport measurements, and hydrostatic pressure tuning. We provide evidence for unconventional superconductivity in 6R-TaS$_{2}$, characterized by an extremely dilute superfluid density, nodal pairing at ambient pressure, and a strong pressure-induced enhancement of superfluid density up to 2 GPa, accompanied by a transition from nodal to nodeless pairing. This evolution correlates with a 50${\%}$ reduction in the CDW onset, a 50${\%}$ suppression of MR below $T^{*}$, and the complete suppression of AHE at $T^{*}$. These findings highlight a direct competition between superconductivity, charge order, and the hidden state below $T^{*}$, impacting both c-axis coherence and the availability of electronic states for pairing. Additionally, our results rule out a magnetic origin for the state below $T^{*}$, suggesting that it arises instead from an intrinsic band structure effect.\\


Figures 1b and 1c show the temperature dependence of resistivity measured under various magnetic fields applied perpendicular and parallel to the c-axis, respectively. The significant difference in the critical magnetic field required to suppress the superconducting transition in these two orientations highlights the quasi-2D nature of superconductivity in 6R-TaS$_{2}$. Above $T_{\rm c}$, three characteristic temperatures ($T^{*}$, $T_{\rm CDW,1}$, $T_{\rm CDW,2}$) are observed, which become evident in the first derivative of resistivity (Fig. 1d). The anomalies at $T_{\rm CDW,2}$ ${\simeq}$ 300 K and $T_{\rm CDW,1}$ ${\simeq}$ 310 K correspond to the previously identified CDW transitions \cite{achari2022alternating}. Specifically, $T_{\rm CDW,2}$ marks the transition to a nearly commensurate CDW phase, while $T_{\rm CDW,1}$ corresponds to the transition into a fully commensurate CDW state. Additionally, a clear anomaly at $T^{*}$ suggests an additional phase transition/crossover, supported by the emergence of large magnetoresistance, a sign change in the Hall effect, and the appearance of an anomalous Hall effect (Figs. 1e–h). $T^{*}$ has also been recently identified \cite{liu2024nematic} and attributed to magnetism, but its precise nature remains unclear.\\

	\begin{figure*}[t!]
		\centering
		\includegraphics[width=\linewidth]{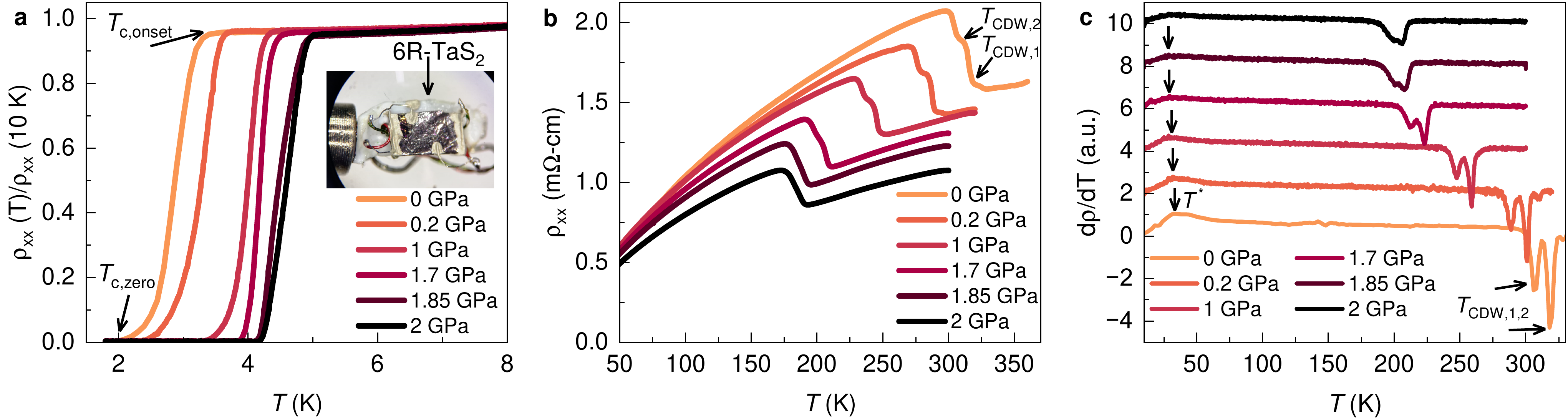}
		\vspace{-0.7cm}
		\caption{\textbf{Resistivity measurements under pressure.}
		$\bf{a,}$ Temperature dependence of resistivity for the single crystal of 6R-TaS$_{2}$, normalized to the value at 8 K, focusing on the low-temperature region to highlight the superconducting transitions recorded at ambient pressure and under hydrostatic pressures up to 2 GPa. $\bf{b-c,}$ Temperature dependence of normal state resistivity and its first derivative for the single crystal of 6R-TaS$_{2}$. Arrows mark three characteristic temperatures $T_{\rm CDW,1}$, $T_{\rm CDW,2}$ and $T^{*}$.}  
		\label{Fig1}
	\end{figure*}

	\begin{figure}[t!]
		\centering
		\includegraphics[width=\linewidth]{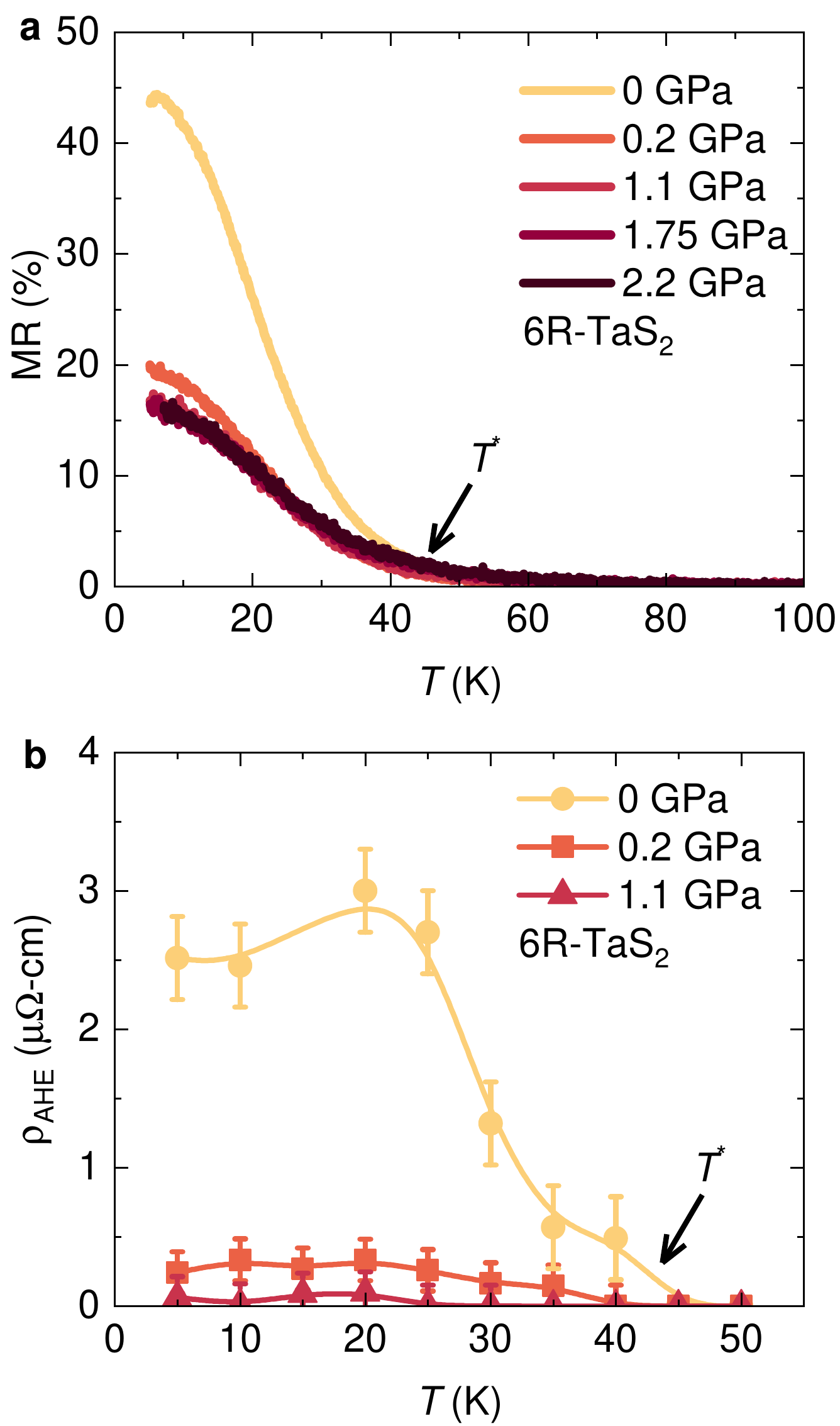}
		\vspace{-0.7cm}
		\caption{\textbf{Magnetoresistance and anomalous Hall resistance under pressure.}
		$\bf{a-b,}$ The temperature dependence of the value of magnetoresistance at 9 T ($\bf{a}$) and the anomalous Hall resistance at 9 T ($\bf{b}$), measured under ambient and various applied hydrostatic pressures.}
		\label{Fig1}
	\end{figure}

Given that multiple transitions are observed in 6R-TaS$_{2}$, we seek to gain deeper insights into both the superconducting and normal state responses using local probe techniques, specifically ${\mu}$SR \cite{amato2024introduction}. A key microscopic property of the superconducting state that can be directly measured using transverse-field ${\mu}$SR (TF-${\mu}$SR) is the superfluid density. This is determined by extracting the second moment of the field distribution from the muon spin depolarization rate ${\sigma}_{{\rm SC}}$, which is related to the superconducting magnetic penetration depth $\lambda$ as $\left<\Delta B^2\right>\propto \sigma^2_{sc} \propto \lambda^{-4}$ (see Methods section). Since $\lambda^{-2}$ is proportional to the superfluid density, ${\sigma}_{{\rm SC}}$ serves as a direct measure. Figure 2a presents TF-${\mu}$SR spectra measured above (3.3~K) and below ($T$ = 80 mK) the superconducting transition temperature $T_{{\rm c}}$. To ensure a well-ordered vortex lattice, measurements were performed after field cooling the sample from above $T_{{\rm c}}$. Above $T_{{\rm c}}$, the oscillations exhibit damping primarily due to random local fields from nuclear magnetic moments, with a nearly constant damping rate down to 2 K. Below $T_{{\rm c}}$, the damping rate increases with decreasing temperature, reflecting the formation of a flux-line lattice in the superconducting state, leading to a nonuniform local magnetic field distribution. Figure 2c shows the temperature dependence of the superconducting relaxation rates $\sigma_{\rm SC-ab}$ and $\sigma_{\rm SC-ac}$, corresponding to the superconducting screening currents with the field applied along the c-axis and within the plane, respectively. The out-of-plane relaxation rate can be extracted as $\sigma_{\rm SC-c}$ = $\sigma_{\rm SC-ac}^{2}$/$\sigma_{\rm SC-ab}$. Notably, the superconducting transition temperature determined from $\mu$SR  is lower than the onset temperature extracted from resistivity measurements, indicating that true bulk 3D superconductivity with a well-arranged vortex state is only established below 2 K.
Since the applied magnetic field is much smaller than the upper critical field ($H_{\rm c2}$), we can use Eq. 2 (see the methods section) to estimate both the in-plane ($\lambda_{ab}$) and out-of-plane ($\lambda_{c}$) penetration depths (see right axis of Fig. 2b). At 0.3 K, we determine: $\lambda_{ab}$ $\simeq$ 733 nm, based on $\sigma_{\rm SC-ab}$, which probes screening currents flowing parallel to the kagome plane. $\lambda_{c}$ $\simeq$ 1857 nm, based on $\sigma_{\rm SC-ac}$ and $\sigma_{\rm SC-ab}$, which probes screening currents flowing perpendicular to the kagome plane. The temperature dependence of the penetration depth provides insights into the superconducting gap topology. In a fully gapped superconductor, $\Delta\lambda^{-2}\left(T\right)\equiv\lambda^{-2}\left(0\right)-\lambda^{-2}\left(T\right)$ vanishes exponentially at low temperatures. In contrast, in a nodal superconductor, it follows a power-law dependence on $T$. To model the gap function, we consider two possibilities: A constant (fully gapped) nodeless order parameter, $\Delta_{0,i}=\Delta_{i}$, and a nodal A d-wave gap, $\Delta_{0,i}=\Delta_{i}\cos2\varphi$, where $\varphi$ is the polar angle around the Fermi surface. The observed behavior of $\lambda^{-2}_{ab}(T)$ is well described by a nodal gap, strongly suggesting nodal superconductivity in 6R-TaS$_{2}$. The combination of an extremely low superfluid density and nodal pairing provides compelling evidence for unconventional superconductivity 6R-TaS$_{2}$. We furthermore note that the data along the $c$-axis differs from those in the $ab$ planes, as $\sigma_{\rm SC,ac}$ exhibits a more saturated-like behavior, consistent with a nodeless gap. A similar distinction between $\sigma_{\rm SC,ab}$ and $\sigma_{\rm SC,ac}$ is also observed in cuprates \cite{PhysRevLett.99.237601}. While the origin of this behavior remains unclear, it highlights a potential similarity between 6R-TaS$_{2}$ and cuprates.\\

	\begin{figure*}[t!]
		\centering
		\includegraphics[width=\linewidth]{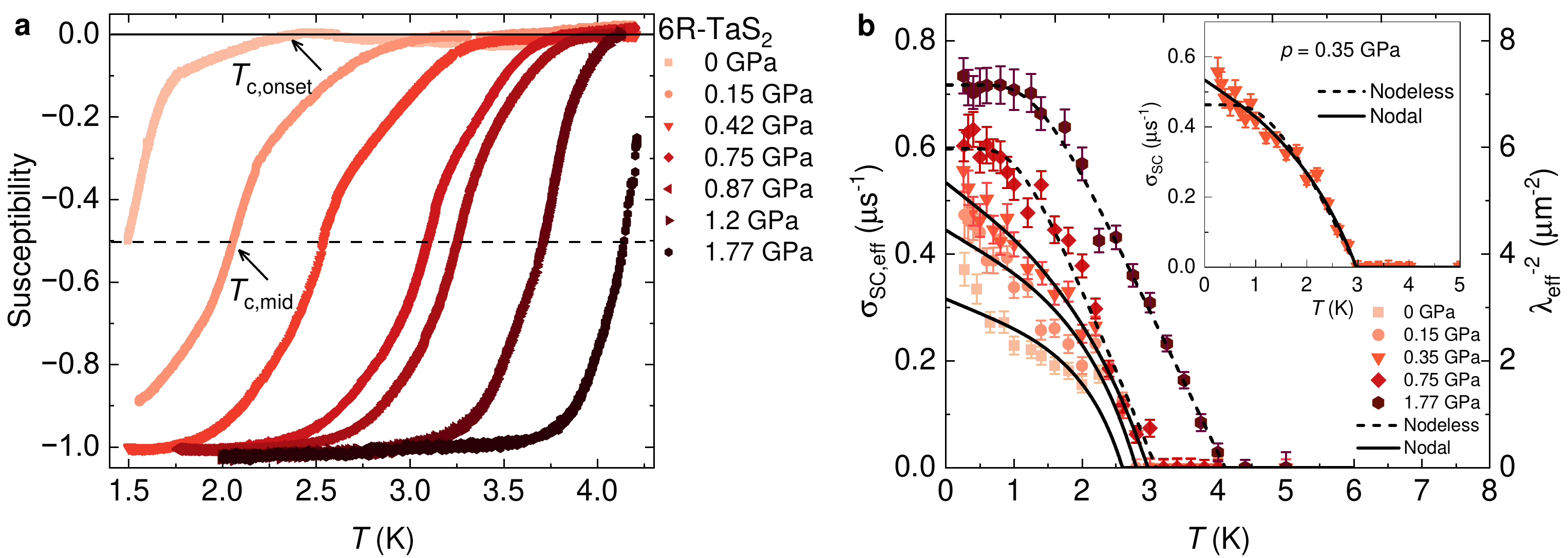}
		\vspace{-0.7cm}
		\caption{\textbf{Magnetic Susceptibility and superfluid density under pressure.}
		$\bf{a,}$ The temperature dependence of the (dia)magnetic susceptibility for the polycrystalline sample of 6R-TaS$_{2}$  recorded at ambient and under various hydrostatic pressures. Arrows mark the onset temperature $T_{\rm c,onset}$ and the temperature $T_{\rm c,mid}$ at which  the susceptibility reaches 50${\%}$. $\bf{b,}$ The temperature dependence of the superconducting muon spin depolarization rate ${\sigma}_{\rm SC,eff}$ for the polycrystalline sample of 6R-TaS$_{2}$, measured in an applied magnetic field of 
        10 mT at ambient and various applied hydrostatic pressures. The error bars represent the standard deviations of the fit parameters. The solid (dashed) lines correspond to a fit using a model with nodeless (nodal) gap superconductivity. Right axis shows the inverse squared effective penetration depth $\lambda_{eff}^{-2}$.}
		\label{Fig1}
	\end{figure*}

To investigate the presence of magnetism in the normal state, particularly below $T^{*}$, we perform ZF-${\mu}$SR experiments. Figure 2b presents the ZF-${\mu}$SR spectra collected over a wide temperature range, revealing a temperature-dependent muon spin relaxation. Since the zero-field relaxation is suppressed upon applying a small longitudinal field ($B_{{\rm LF}}$~=~5~mT), the observed relaxation originates from static spontaneous fields, which persist on the microsecond timescale. 
To analyze the data, we fit the ZF-${\mu}$SR spectra using a stretched Gaussian Kubo-Toyabe depolarization function\cite{kubo1967magnetic,musrfit}, which provides an intermediate fitting model between Gaussian and Lorentzian relaxation. This approach assumes a static and randomly distributed local field:

\begin{equation}
\begin{aligned}
A_{ZF}^{GKT}(t) =  A_{0}\Big[\frac{1}{3} + \frac{2}{3}\Big(1 - (\Gamma t)^{\beta}\Big) \exp\Big[-\frac{(\Gamma t)^{\beta}}{\beta}\Big]\Big]\\ 
\end{aligned}
\end{equation}  

In our analysis, the parameter ${\beta}$ varies between 1 and 2. While ${\beta}$ continuously interpolates between a purely Gaussian (${\beta}$=2) and a purely Lorentzian (${\beta}$=1) static Kubo-Toyabe lineshape, its physical significance remains unclear. The parameter ${\Gamma}$ in Eq. 1 represents the phenomenological muon-spin relaxation rate, which reflects the width of the local field distribution. Since this function smoothly transitions between Gaussian and Lorentzian lineshapes, the absolute value of ${\Gamma}$ cannot be directly interpreted in cases where ${\beta}$ is variable. Figure 2d presents the temperature dependence of ${\beta}$ and ${\Gamma}$. Above $T_{\rm CDW}$, ${\beta}$ maintains a value of 2, but it gradually decreases below $T_{\rm CDW}$, reaching ${\beta}$=1 at 250 K, where it remains nearly constant down to 5 K. In contrast, ${\Gamma}$ exhibits a non-monotonic behavior, reaching a minimum around 250 K, followed by a gradual increase and eventual saturation below 100 K. The significant evolution of the ZF-${\mu}$SR signal lineshape suggests a strong coupling between muons and the electronic environment. The onset of CDW order likely modifies the electric field gradient, altering the coupling between muons and nuclear spins. The gradual increase in ${\Gamma}$ below 250 K may reflect a continuous redistribution of charge carriers, eventually stabilizing below 100 K. Crucially, the absence of any anomaly in ${\beta}$ or an increase in ${\Gamma}$ below $T^{*}$ indicates that no spontaneous magnetism emerges at $T^{*}$ in 6R-TaS$_{2}$. This strongly suggests that the large AHE and MR observed below $T^{*}$ are not linked to magnetism but instead originate from band structure effects, likely of topological origin.\\
   
In the following, we explore how superconductivity, charge order, and the hidden order in 6R-TaS$_{2}$ evolve under hydrostatic pressure. The resistivity vs. temperature curves in Figure 3 reveal a broad superconducting transition, spanning approximately 1 K, with two characteristic temperatures: $T_{\rm c,onset}$, marking the onset of superconducting correlations, and $T_{\rm c,zero}$, below which resistance drops to zero. We attribute this broad transition to $c$-axis frustration of superconducting coupling due to the presence of charge density wave (CDW) order. In this scenario, $T_{\rm c,onset}$ represents the temperature where superconducting correlations first develop within the planes, while 3D superconductivity is established only below $T_{\rm c,zero}$. This interpretation is further supported by ${\mu}$SR results, which detect the superconducting transition only below $T_{\rm c,zero}$. Upon applying hydrostatic pressure up to 2 GPa, both $T_{\rm c,onset}$ and $T_{\rm c,zero}$ increase by a factor of two, while the difference between them is reduced, leading to a sharper superconducting transition. At the same time, the onset of CDW order undergoes substantial suppression, shifting from $T_{\rm CDW,1}$ ${\simeq}$ 310 K to $T_{\rm CDW,1}$ ${\simeq}$ 180 K. Furthermore, the two-step anomaly associated with the CDW transition merges into a single-step anomaly, indicating that pressure suppresses the intermediate transition from a nearly commensurate to a fully commensurate CDW state. The observed weakening of CDW order and the corresponding enhancement of superconductivity, along with the sharpening of the superconducting transition, strongly suggest that CDW order frustrates superconducting coupling, and its suppression under pressure enables the emergence of a more robust superconducting state. This is substantiated by the diamagnetic susceptibility measurements on polycrystalline 6R-TaS$_{2}$ (see Fig. 5a). At ambient pressure, a two-step superconducting transition is observed, consistent with the evolution of superconductivity from a 2D to a 3D state. To characterize the changes in the superconducting critical temperature, we identify $T_{\rm c,onset}$ as the onset temperature and $T_{\rm c,mid}$-as a reliable indicator of 3D superconducting order. Under pressure, the two-step transition merges into a single, sharp superconducting transition, while $T_{\rm c}$ increases to nearly 5 K, aligning well with the single-crystal results presented above. These findings further confirm the crucial role of pressure in enhancing superconductivity while suppressing competing charge order in 6R-TaS$_{2}$.\\ 

	\begin{figure*}[t!]
		\centering
		\includegraphics[width=\linewidth]{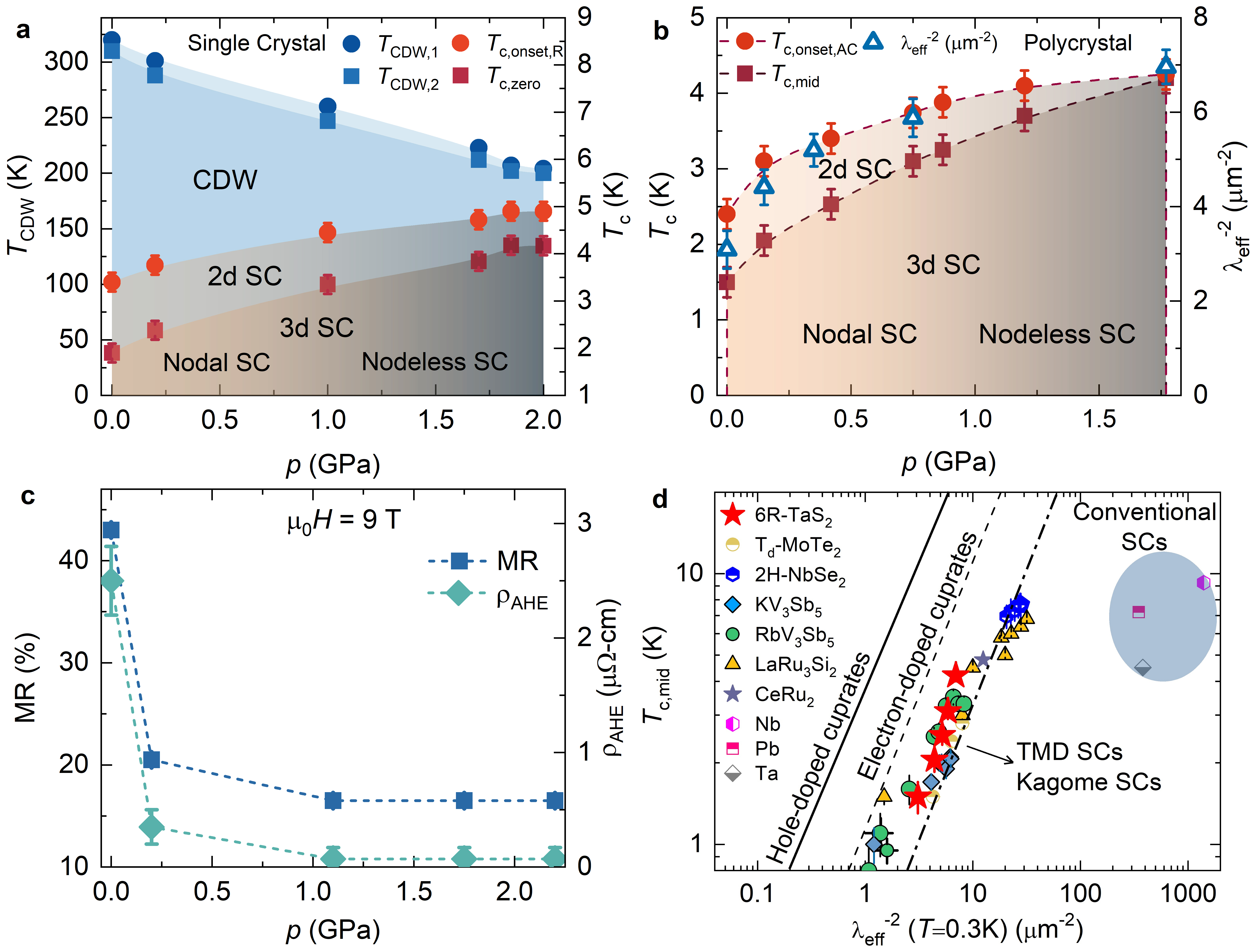}
		\vspace{-0.7cm}
		\caption{\textbf{Phase Diagrams.}
		$\bf{a,}$ Pressure dependence of the charge density wave transition temperatures $T_{\rm CDW,1}$ and $T_{\rm CDW,2}$, the superconducting transition temperatures $T_{\rm c,onset,R}$ and $T_{\rm c,zero}$, determined from the resistivity experiments on the single crystal sample of 6R-TaS$_{2}$. $\bf{b,}$ Pressure dependence of the superconducting transition temperatures $T_{\rm c,onset,AC}$ and $T_{\rm c,mid}$, determined from AC susceptibility measurements, and the zero-temperature value of the inverse squared effective penetration depth $\lambda_{eff}^{-2}$. $\bf{c,}$ Pressure dependence of MR and  $\rho_{\rm AHE}$, recorded at 5 K and under 9 T.  $\bf{d,}$ Plot of $T_{\rm c,mid}$ versus $\lambda_{eff}^{-2}$ on a logarithmic scale obtained from ${\mu}$SR experiments for the various kagome-lattice superconductors \cite{guguchia2023unconventional}. The dash-dotted line represents the relationship obtained for the layered transition metal dichalcogenide superconductors \cite{guguchia2017signatures,doi:10.1126/sciadv.aav8465}. The relationship observed for cuprates is also shown \cite{Uemura1,Shengelaya} as well as the points for various conventional superconductors.}
		\label{Fig1}
	\end{figure*}

Additionally, the anomaly at $T^{*}$, associated with hidden order and anomalous transport properties, gradually weakens under pressure, while its temperature remains unchanged. This suggests that pressure negatively impacts MR and AHE below $T^{*}$. To explore this further, we measured both quantities as a function of pressure. Figure 4 shows the temperature dependence of MR under ambient conditions with various applied pressures up to $p$=2 GPa. Notably, while the onset temperature $T^{*}$ remains nearly unchanged, the absolute MR value decreases significantly from 42${\%}$ to 21${\%}$ within 0.2 GPa; beyond which it only reduces slightly before saturating at 18${\%}$. Additionally, the AHE is suppressed by a factor of 8 within 0.2 GPa, above which it continues to decrease and becomes fully suppressed within 1 GPa, or is reduced to a level undetectable within our experimental precision. These findings indicate that the pressure-induced enhancement of superconductivity is accompanied by a substantial weakening of not only the charge order onset temperatures $T_{\rm CDW,1,2}$ but also the MR and the AHE linked to $T^{*}$.\\
 
The above findings underscore a strong competition between superconductivity, charge order, and hidden order in 6R-TaS$_{2}$, with pressure serving as a key tuning parameter to modify their interplay. An important question arises: does this competition emerge solely from interlayer frustration, or do these orders also compete for the same electronic states? To address this, we examine the effect of pressure on superfluid density, which provides a direct measure of the number of Cooper pairs. Figure 5b summarizes the pressure and temperature dependence of ${\sigma}_{{\rm SC,eff}}$ (measured in an applied magnetic field of ${\mu}_{\rm 0}H = 10$~mT) for polycrystalline 6R-TaS$_{2}$. From ${\sigma}_{{\rm sc,eff}}$, we extract the magnetic penetration depth $\lambda_{\rm eff}$ (see Methods), plotted on the right axis of Figure 5b. For polycrystalline samples, this yields an effective penetration depth $\lambda_{\rm eff}$. 
As the temperature decreases below $T_{{\rm c}}$, the depolarization rate ${\sigma}_{{\rm sc,eff}}$ increases from zero, indicating the formation of a flux-line lattice (Fig. 5b). With increasing pressure, not only does $T_{{\rm c}}$ (determined from AC susceptibility and ${\mu}$SR experiments) rise, but the low-temperature value of ${\sigma}_{{\rm SC}}$ (measured at 0.25 K) also increases by at least a factor of two within 2 GPa. This demonstrates that superconductivity and charge order compete for the same electronic states, and when charge order is suppressed, these states become available, leading to an enhancement in superfluid density. To further probe the nature of superconductivity, we analyze the temperature dependence of the penetration depth $\lambda_{\rm eff}$. Generally, a power-law dependence of $\lambda_{\rm eff}^{-2}(T)$ at low temperatures indicates nodal quasiparticles, whereas exponential saturation is characteristic of a fully gapped superconducting state. In polycrystalline 6R-TaS$_{2}$, the low-temperature behavior of $\lambda_{ab}^{-2}(T)$, measured down to 300 mK, exhibits linear $T$ dependence under pressures of $p$= 0 GPa, 0.15 GPa and 0.35 GPa, consistent with the presence of gap nodes. This aligns with the nodal behavior observed in the single-crystal sample of 6R-TaS$_{2}$ at ambient pressure. However, for higher pressures ($p$=0.75 GPa and $p$=1.77 GPa), $\lambda_{\rm eff}^{-2}(T)$ transitions to a saturation-like behavior, which is well described by a nodeless superconducting gap model. A quantitative ${\chi_r}^{2}$ comparison confirms this transition: at $p$=0.75 GPa and $p$=1.77 GPa, the nodal gap model yields a ${\chi_r}^{2}$ value 3.9 times larger than that of the nodeless gap model, indicating a significantly better fit for the latter. In contrast, for pressures below 0.75 GPa, the nodal gap model provides a better fit than the nodeless one. These experiments demonstrate that CDW order not only competes with superconductivity for the same electronic states but also modifies the superconducting gap structure, driving a transition from nodal to nodeless pairing under pressure.\\

To summarize, understanding the interplay between quantum phases—such as superconductivity, charge order, magnetism, and anomalous transport—is a central topic in condensed matter physics, particularly for unraveling the mechanisms behind unconventional superconductivity. The transition metal dichalcogenide system 6R-TaS$_{2}$ offers a rare platform where this rich physics can be explored using external tuning and a multi-technique approach. In this work, we report several key findings, summarised in Figures 6a-d:
(1) We reproduce charge order transitions below $T_{\rm CDW,1,2}$, observed as a strong anomaly in electrical resistivity, and hidden order below  $T^{*}$, manifested by the appearance of magnetoresistance and the anomalous Hall effect, as well as the onset of superconductivity below $T_{\rm c,onset,R}$ ${\simeq}$ 3.5 K. 
(2) No spontaneous magnetic order is detected below $T^{*}$, suggesting that AHE arises from topological band structure effects rather than magnetism.
(3) The onset of CDW order is suppressed by a factor of two upon application of hydrostatic pressure of 2 GPa (Fig. 6a).
(4) The most striking effect is observed in MR and AHE (Fig. 6c). Namely, the absolute value of magnetoresistance, linked to hidden order, is reduced by a factor of two at an extremely low pressure of 0.2 GPa, above which it remains nearly unchanged. Furthermore, the anomalous Hall effect is completely suppressed within 0.2 GPa, demonstrating the extreme sensitivity of AHE to pressure in 6R-TaS$_{2}$. 
(5) At ambient pressure, we uncover a nodal superconducting state with an exceptionally low superfluid density. This nodal structure may not be symmetry-induced; however, one possible explanation for this result is the changes induced by the emergence of charge order on the Fermi surface \cite{guguchia2023tunable,fernandes2010transfer}. The onset of long-range charge order fragments the pockets into additional smaller pockets. Depending on the details of the reconstructed Fermi pockets, accidental nodes could emerge.
(6) At ambient pressure, the superconducting transition is broad, as most clearly evident in magnetic susceptibility. The onset likely corresponds to the development of 2D superconductivity within the layers, with coherent 3D superconductivity emerging only at lower temperatures (Fig. 6a and b).
(7) Under pressures up to 2 GPa, the superconducting transition temperature and superfluid density increase significantly, the superconducting transition sharpens and the superconducting pairing structure transitions from nodal to nodeless (Fig. 6a and b).
(8) We observed a strong correlation between $T_{\rm c,mid}$ and $\lambda_{eff}^{-2}$, as shown in Fig. 6d. The data points for 6R-TaS$_{2}$ fall along the same trend line as other TMDs and kagome superconductors, and are positioned close to the cuprates \cite{Uemura1,Shengelaya}. Such a correlation is a hallmark of unconventional superconductivity \cite{Uemura1,emery1995importance}. These findings imply unconventional superconductivity in 6R-TaS$_{2}$. Furthermore, these results highlight the delicate balance between electronic interactions in 6R-TaS$_{2}$ and imply that pressure serves as an effective tuning parameter for manipulating competing quantum phases. The transition from nodal to nodeless superconductivity, alongside the suppression of CDW and AHE, suggests a strong coupling between these orders and raises intriguing questions about the underlying pairing mechanism and the role of intertwined electronic correlations.  We note that the strong interdependence between superconductivity, charge order, and hidden order, as manifested by MR and AHE, is markedly different from observations in 2H/4H-NbSe$_{2}$ and 2H-NbS$_{2}$ \cite{doi:10.1126/sciadv.aav8465,islam2024pressure}. In the latter systems, pressure substantially modifies the superfluid density, while its effect on charge order or MR remains modest.\\

This result has broader implications, as such a complex phase diagram is a hallmark of a wider class of unconventional superconductors, including cuprates, iron-based superconductors, kagome-lattice systems, and other correlated electron systems. The observed tunability of superconductivity and charge order under pressure underscores the importance of external control parameters in revealing hidden phases and novel electronic states. This work not only deepens our understanding of the interplay between competing orders in 6R-TaS$_{2}$ but may also provide valuable insights into the mechanisms governing unconventional superconductivity in strongly correlated materials.\\

\section{Acknowledgments}~
The ${\mu}$SR experiments were carried out at the Swiss Muon Source (S${\mu}$S) Paul Scherrer Insitute, Villigen, Switzerland. Z.G. acknowledges support from the Swiss National Science Foundation (SNSF) through SNSF Starting Grant (No. TMSGI2${\_}$211750).\\ 

\section{Author contributions}~
Z.G. conceived and supervised the project. Growth of single crystals: A.A. and R.R.N.. Growth of Polycrystalline sample: J.N.T. and F.v.R.. 
Magnetotransport experiments under pressure: V.S., P.K., O.G. and Z.G.. AC susceptibility experiments under pressure: Z.G., J.N.G., S.S.I., V.S..  
$\mu$SR experiments under pressure, the corresponding analysis and discussions: V.S., J.N.G., S.S.I., O.G., P.K., H.G., G.S., M.J., M.B., R.K., H.L., and Z.G.. Figure development and writing of the paper: Z.G. and V.S. All authors discussed the results, interpretation, and conclusion.\\

\section{Methods}

\textbf{Sample preparation}: Single crystal samples of 6R-TaS$_{2}$ are prepared by the procedure reported in Ref. \cite{achari2022alternating}.\\

\textbf{${\mu}$SR experiment}: In a ${\mu}$SR experiment \cite{amato2024introduction} nearly 100 ${\%}$ spin-polarized muons ${\mu}$$^{+}$
are implanted into the sample one at a time. The positively charged ${\mu}$$^{+}$ thermalize at interstitial lattice sites, where they
act as magnetic microprobes. In a magnetic material the muon spin precesses in the local field $B_{\rm \mu}$ at the
muon site with the Larmor frequency ${\nu}_{\rm \mu}$ = $\gamma_{\rm \mu}$/(2${\pi})$$B_{\rm \mu}$ (muon
gyromagnetic ratio $\gamma_{\rm \mu}$/(2${\pi}$) = 135.5 MHz T$^{-1}$). Using the $\mu$SR technique, important length scales of superconductors can be measured, namely the magnetic penetration depth $\lambda$ and the coherence length $\xi$. If a type II superconductor is cooled below $T_{\rm c}$ in an applied magnetic field ranging between the lower ($H_{c1}$) and the upper ($H_{c2}$) critical fields, a vortex lattice is formed which in general is incommensurate with the crystal lattice, with vortex cores separated by much larger distances than those of the crystallographic unit cell. Because the implanted muons stop at given crystallographic sites, they will randomly probe the field distribution of the vortex lattice. Such measurements need to be performed in a field applied perpendicular to the initial muon spin polarization (so-called transrverse-field, TF configuration).\\

The magnetic penetration depth ${\lambda}(T)$ is related to the superconducting muon spin depolarization rate 
${\sigma}_{{\rm SC}}(T)$ in the presence of a triangular (or hexagonal) vortex lattice by the equation \cite{amato2024introduction,mielke2022time}: 
\begin{equation}
\frac{\sigma_{SC}(T)}{\gamma_{\mu}}=0.06091\frac{\Phi_{0}}{\lambda^{2}(T)},
\end{equation}
where ${\gamma_{\mu}}$ is the gyromagnetic ratio of the muon and ${\Phi}_{{\rm 0}}$ is the magnetic-flux quantum. 
Since the applied field is a factor of ${\sim}$ 30 times smaller than the second critical magnetic fields in 6R-TaS$_{2}$, this equation is valid to estimate the $\lambda$.\\

Pressures up to 1.9 GPa were generated in a double wall piston-cylinder type cell made of CuBe/MP35N, specially designed to perform ${\mu}$SR experiments under pressure \cite{khasanov2016high}. As a pressure transmitting medium Daphne oil was used. The pressure was measured by tracking the superconducting transition of a very small indium plate by AC susceptibility. The filling factor of the pressure cell was maximized. The fraction of the muons stopping in the sample was approximately 40 ${\%}$.\\

\textbf{Resistivity experiments}:
Experiments under ambient pressure were performed by using 
the DC resistance bridge of the Quantum Design Physical Property Measurement System. Experiments under pressure were performed with the same instrument by using double wall piston cylinder type of cell from Almax easylab Pcell15/30 module. The sample was mounted in the clamp cell such that the magnetic field was applied perpendicular to the in-plane surface of the sample. As a pressure transmitting medium Daphne oil was used. Pressure at room temperature is monitored by measuring the resistance of a Manganin wire, enabling precise pressure control. The pressure at low temperature was measured by tracking the superconducting transition of Tin (Sn) (see Fig. 9). For Magnetoresistance (MR) and Hall resistivity measurements, standard symmetrization and antisymmetrization procedures were applied to the signal, respectively, to accurately extract the desired information.\\

\textbf{Analysis of TF-${\mu}$SR data under pressure}: The TF ${\mu}$SR data were analyzed by using the following functional form:\cite{guguchia2023tunable}.
\begin{equation}
\begin{aligned}
P(t)=A_s\exp\Big[-\frac{(\sigma_{\rm sc}^2+\sigma_{\rm nm}^2)t^2}{2}\Big]\cos(\gamma_{\mu}B_{\rm int,s}t+\varphi) \\
 + A_{\rm pc}\exp\Big[-\frac{\sigma_{\rm pc}^2t^2}{2}\Big]\cos(\gamma_{\mu}B_{\rm int,pc}t+\varphi), 
\end{aligned}
\end{equation}
 Here $A_{\rm s}$ and $A_{\rm pc}$  denote the initial asymmetries of the sample and the pressure cell, respectively. ${\varphi}$ is the initial phase of the muon-spin ensemble and $B_{\rm int}$ represents the internal magnetic field at the muon site. The relaxation rates ${\sigma}_{\rm sc}$ and ${\sigma}_{\rm nm}$ characterize the damping due to the formation of the FLL in the superconducting state and of the nuclear magnetic dipolar contribution, respectively. In the analysis ${\sigma}_{\rm nm}$ was assumed to be constant over the entire temperature range and was fixed to the value obtained above $T_{\rm c}$ where only nuclear magnetic moments contribute to the muon depolarization rate ${\sigma}$. The Gaussian relaxation rate, ${\sigma}_{\rm pc}$, reflects the depolarization due to the nuclear moments of the pressure cell.\\

 \textbf{Analysis of ${\lambda}(T)$}: ${\lambda}$($T$) was calculated within the local (London) approximation (${\lambda}$ ${\gg}$ ${\xi}$) by the following expression \cite{musrfit}: 
\begin{equation}
\frac{\lambda^{-2}(T,\Delta_{0,i})}{\lambda^{-2}(0,\Delta_{0,i})}= 1+\frac{1}{\pi}\int_{0}^{2\pi}\int_{\Delta_i(_{T,\varphi})}^{\infty}(\frac{\partial f}{\partial E})\frac{EdEd\varphi}{\sqrt{E^2-\Delta_i(T,\varphi)^2}},
\end{equation}
where $f=[1+\exp(E/k_{\rm B}T)]^{-1}$ is the Fermi function, ${\varphi}$ is the angle along the Fermi surface, and ${\Delta}_{i}(T,{\varphi})={\Delta}_{0,i}{\Gamma}(T/T_{\rm c})g({\varphi}$)
(${\Delta}_{0,i}$ is the maximum gap value at $T=0$). 
The temperature dependence of the gap is approximated by the expression 
${\Gamma}(T/T_{\rm c})=\tanh{\{}1.82[1.018(T_{\rm c}/T-1)]^{0.51}{\}}$,\cite{carrington2003magnetic} 
while $g({\varphi}$) describes the angular dependence of the gap and it is replaced by 1 for both an $s$-wave and an $s$+$s$-wave gap,
${\mid}\cos(2{\varphi}){\mid}$ for a $d$-wave gap, and ${\mid}\cos(6{\varphi}){\mid}$ for a $f$-wave gap.\\

\bibliography{References}{}

	\begin{figure*}[t!]
		\centering
		\includegraphics[width=\linewidth]{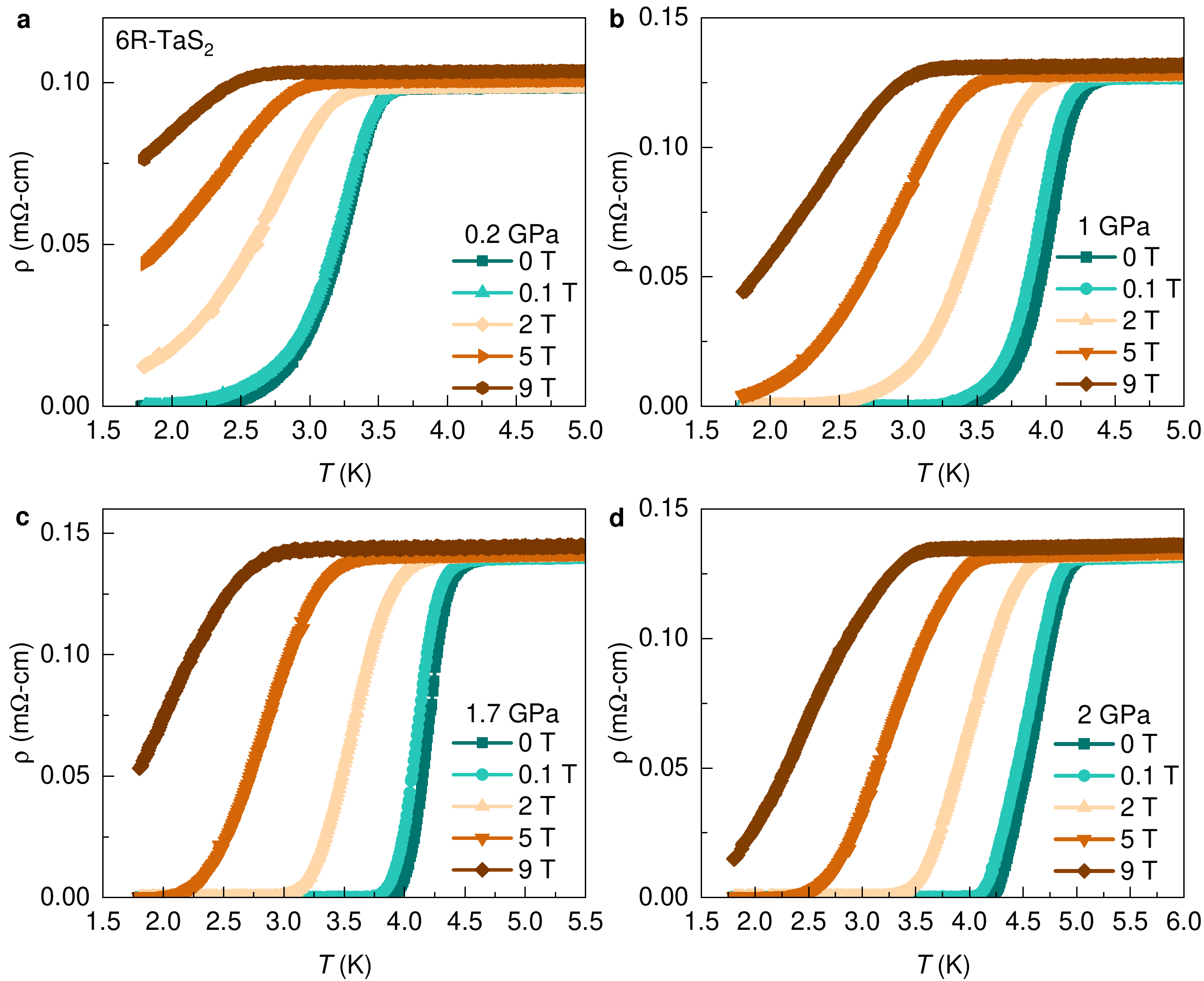}
		\vspace{-0.7cm}
		\caption{\textbf{Field dependent resistivity measurements under various pressure.}
		$\bf{a-d,}$ The temperature dependence of the electrical resistivity of a single crystal of 6R-TaS$_{2}$, focusing on the low-temperature region, measured under various hydrostatic pressures and under various magnetic fields applied perpendicular to the $c$-axis.}
		\label{Fig1}
	\end{figure*}

	\begin{figure*}[t!]
		\centering
		\includegraphics[width=\linewidth]{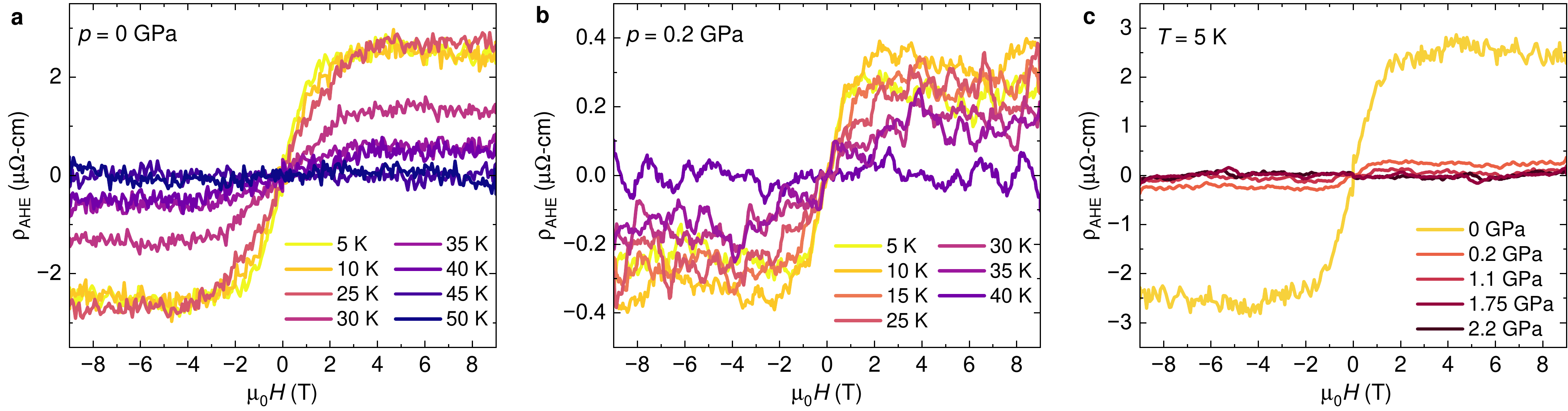}
		\vspace{-0.7cm}
		\caption{\textbf{Pressure tuning of AHE.}
		$\bf{a-b,}$ Field dependence of anomalous Hall resistance $\rho_{\rm AHE}$ at $p$ = 0 GPa ($\bf{a}$) and 0.2 GPa ($\bf{b}$), measured at various temepratures. $\bf{c,}$ Field dependence of anomalous Hall resistance $\rho_{\rm AHE}$, recorded at 5 K for various hydrostatic pressure conditions.}
		\label{Fig1}
	\end{figure*}

	\begin{figure*}[t!]
		\centering
		\includegraphics[width=0.6\linewidth]{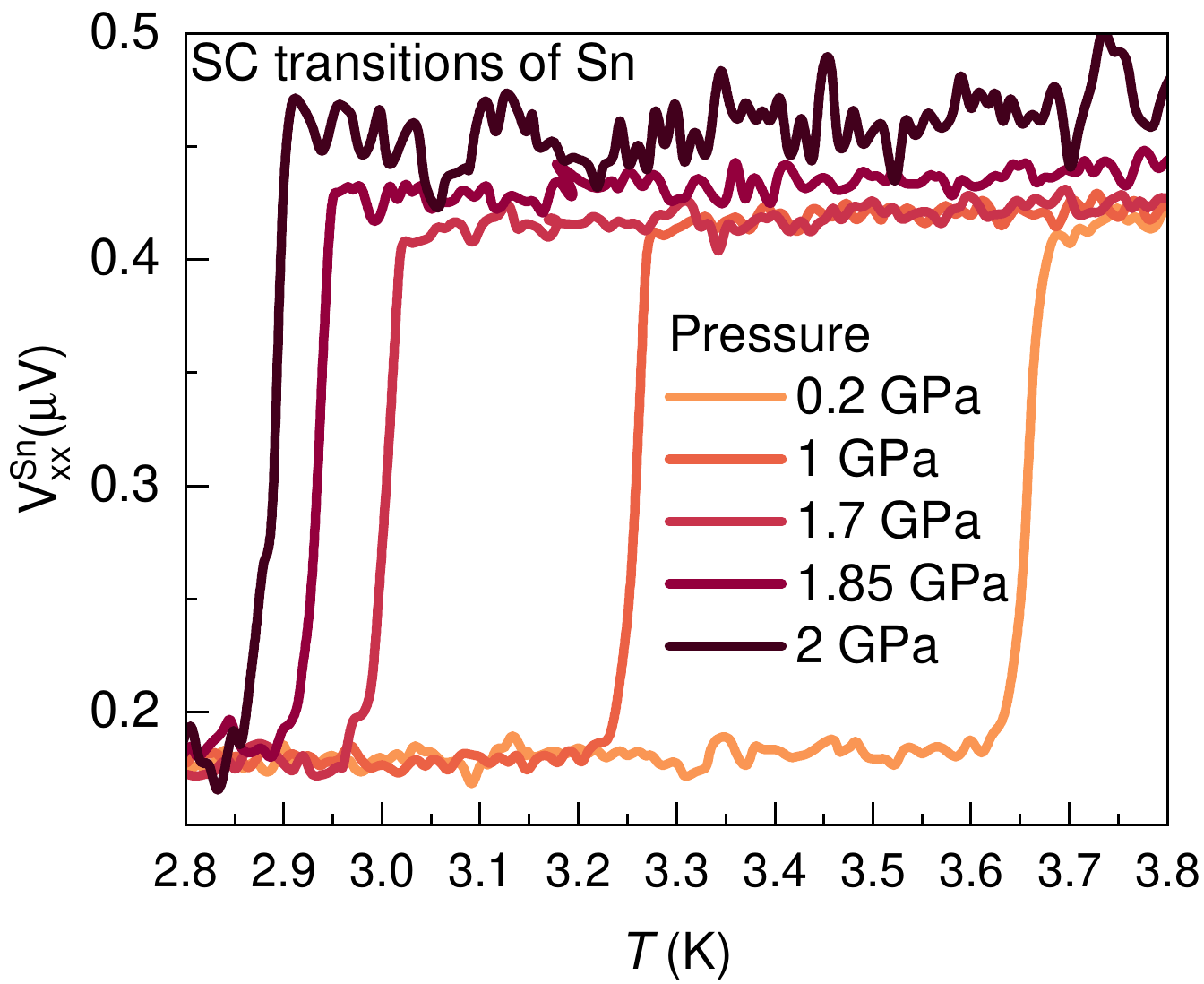}
		\vspace{0cm}
		\caption{\textbf{Superconducting transition of Sn as a pressure indicatior.}
		The superconducting transition of Sn under various applied pressures, used to estimate the pressure in the pressure cell.}
		\label{Fig1}
	\end{figure*}

\end{document}